\documentclass[sigconf]{acmart}

\usepackage{balance}
\AtBeginDocument{%
  }

\setcopyright{acmlicensed}
\copyrightyear{2018}
\acmYear{2018}
\acmDOI{https://doi.org/10.1145/3803437.3805581}
\acmConference[Conference acronym 'XX]{Make sure to enter the correct
  conference title from your rights confirmation email}{June 03--05,
  2018}{Woodstock, NY}
\acmISBN{978-1-4503-XXXX-X/2018/06}

\usepackage{xcolor}

\copyrightyear{2026}
\acmYear{2026}
\setcopyright{cc}
\setcctype{by}
\acmConference[FSE Companion '26]{34th ACM Joint European Software Engineering Conference and Symposium on the Foundations of Software Engineering}{July 05--09, 2026}{Montreal, QC, Canada}
\acmBooktitle{34th ACM Joint European Software Engineering Conference and Symposium on the Foundations of Software Engineering (FSE Companion '26), July 05--09, 2026, Montreal, QC, Canada}
\acmDOI{10.1145/3803437.3805581}
\acmISBN{979-8-4007-2636-1/2026/07}
\begin{document}

\title{At What Cost? Software Developers' Well-Being in the Age of GenAI}

\author{Mariam Guizani}
\email{mariam.guizani@queensu.ca}
\affiliation{\institution{Queen's University}
\country{Canada}}

\author{Maduka Subasinghage}
\email{maduka.subasinghage@uwa.edu.au}
\affiliation{\institution{The University of Western Australia}
\country{Australia}}

\author{Sherlock A. Licorish}
\email{sherlock.licorish@otago.ac.nz}
\affiliation{\institution{University of Otago}
\country{New Zealand}}

\author{Sofia Ouhbi}
\email{sofia.ouhbi@it.uu.se}
\affiliation{\institution{Uppsala University}
\country{Sweden}}


\renewcommand{\shortauthors}{}

\begin{abstract}
  Generative Artificial Intelligence (GenAI) is rapidly reshaping software development, with growing emphasis on accelerating productivity and optimizing performance. However, excessive focus on such dimensions risks overlooking the critical implications for developer well-being. GenAI tools can amplify cognitive load, introduce new forms of oversight labor, and escalate expectations
around output and pace, contributing to stress, burnout, and diminished work–life balance. The GenAI movement is also transforming professional norms, altering career entry points, demanding
continuous adaptation, and deepening inequalities in access and support. This position paper calls for a reorientation of the GenAI research agenda in software development and proposes a theoretical framework to move beyond narrow performance metrics toward investigations that also center on human experience, social context, and sustainable productivity. 
\end{abstract}




\maketitle
\section{Introduction}
Generative artificial intelligence (GenAI) tools, such as large language models, are rapidly reshaping software development processes. Empirical studies in software engineering have investigated developers' experience using GenAI in terms of short-term productivity gains, time to complete tasks, code quality, and overall perceived usefulness \cite{11190120, licorish2025comparinghumanllmgenerated, mohamed2025impact}. Although outcomes are mixed, GenAI is increasingly positioned as a solution to long-standing efficiency challenges in software development. 

Despite the rapid growth in GenAI technologies, there is limited understanding of how they affect the well-being of software developers, particularly those required to integrate them into their workflows under the assumption that productivity gains will follow. The current research has been mainly focused on productivity and efficiency concerns, for instance, answering questions such as ``Can GenAI perform X?'' \cite{11190120} and comparing GenAI's outputs' quality and correctness to that of developers \cite{licorish2025comparinghumanllmgenerated}. In a recent systematic literature review, Mohamed et al. \cite{mohamed2025impact} showed that most studies are focused on performance outcomes, such as speed, correctness, or code quality, and largely overlook human outcomes beyond usability and intent to adopt GenAI. Some of these performance outcomes are contested areas with evidence pointing to both improvements and degradations depending on
context \cite{mohamed2025impact}. This mirrors earlier trends in developer tooling research, where success is often evaluated in terms of output rather than sustained human experience.

Given the clear implications of GenAI for developers’ work practices, it is critical to examine its effects on well-being more systematically. Well-being encompasses not only job satisfaction, but also cognitive load, emotional exhaustion, burnout, anxiety, and work–life balance \cite{pandey2025systematic, sarkar2024mental}. These aspects highly influence one's quality of life (i.e., health, prosperity, society, environment, and good governance) \cite{GovCanadaQualityLifeFramework2025}. This tension aligns with long-established findings outside software development. Research in organizational psychology and human–computer interaction has shown that technology that enhances productivity often leads to work intensification, where performance standards rise in response to efficiency gains rather than remaining constant \cite{chesley2014information, mazmanian2013autonomy, tarafdar2007impact}. To date, only a small number of studies explicitly theorize or measure outcomes such as burnout, stress, or work–life balance in the context of AI-assisted development \cite{feng2025gains}. This omission is concerning given the pace and scale of GenAI adoption, and the central role developers play in continuously adapting to rapidly evolving tools. 

Studying these effects is crucial for designing responsible AI adoption strategies that safeguard developers’ well-being while enabling organizations to benefit from GenAI‑enhanced workflows. This position paper articulates why developer well-being warrants focused attention in the age of GenAI and proposes a framework for studying software developers’ well-being. We argue for the reorientation of the GenAI research agenda in software engineering, moving beyond narrow performance metrics toward frameworks that also center on human experience, social context, and sustainable productivity.

This paper is positioned as a software engineering research agenda that extends current GenAI evaluation practices by drawing on theories from psychology and sociology. These theories serve as analytical lenses to enrich software engineering research. Our goal is to broaden evaluation beyond performance metrics by incorporating human, social, and organizational dimensions of software developers' work.

\section{Practitioners' Perceptions and Early Signals}

Among other professionals, software developers were reported to use GenAI extensively \cite{Singla2025StateAI}. Beyond GenAI use, recent studies are beginning to explore how professionals perceive the impact of GenAI on their roles. An online questionnaire study \cite{malheiros2024impact} conducted in 2024 examined the impact of GenAI technologies on software development professionals in Brazil, with a particular focus on perceptions of job security. Responses from 39 participants indicated that while most professionals do not currently feel threatened by AI tools, there is considerable uncertainty about their long-term implications. The study also highlighted key benefits, including reduced coding time and improved documentation quality. However, challenges such as unclear communication with AI systems and technical limitations were also reported. Another study \cite{kuhail2024will} conducted a questionnaire with 99 programmers to examine perceptions of ChatGPT and other AI tools in software development. Participants reported using AI for tasks such as generating boilerplate code, explaining complex code, and conducting research, with over half indicating increased productivity. Higher productivity was associated with greater trust in AI tools. While most did not perceive an immediate threat to job security, approximately half expressed concern about future impacts, particularly on entry-level roles.

Similarly, research with practitioners, for instance in creative and design fields, highlights both the promise and the current limitations of GenAI. A study \cite{takaffoli2024generative} involving interviews with 24 professionals explored the use of GenAI in user experience design and research. Its findings reveal a lack of formal Generative AI policies within organizations, predominantly individual (rather than team-based) use of GenAI—mainly for writing tasks as opposed to design-oriented activities—and a widespread need for targeted training to enhance prompt engineering and the evaluation of AI-generated outputs. A study \cite{palani2024evolving} involving 10 creative practitioners across 19 domains, based on interviews and video data, found that users perceived GenAI creativity support tools (e.g., Illustrator and AutoCAD) as fundamentally different from non-GenAI tools. Despite facing UX challenges—such as difficulty articulating goals, misaligned outputs, and fragmented tool ecosystems—practitioners valued GenAI for helping overcome creative blocks, refine ideas, streamline workflows, and to rapidly generate alternatives. Similar findings were echoed in a qualitative study \cite{inie2023designing} that investigated how 23 creative professionals perceive GenAI.

Despite growing interest in professionals’ perceptions of how GenAI shapes their work and performance and some positive outcomes reported, a significant research gap persists regarding GenAI effects on practitioners’ well-being. This gap is especially pronounced when accounting for both individual and social dimensions. Addressing this gap is critical, as focusing solely on productivity neglects the equally important aspect of well-being, which is essential for fostering sustainable, healthy work environments.

\section{The GenAI Paradox in Developer Work}

GenAI tools introduce a paradox in the developer experience. They can reduce effort by offloading routine coding tasks and reducing task initiation overhead at times, albeit not for some experienced developers \cite{mohamed2025impact}. Whenever benefits are had, it is suggested that GenAI can act as a job resource that helps developers accomplish work with less strain. However, GenAI tools can also function as a demand amplifier. Developers must expend mental effort to review, evaluate, and correct AI outputs, which imposes an invisible oversight labor. What is saved in coding can be offset by debugging or quality assurance. For instance, Becker et al. found that the use of AI tools resulted in 19\% productivity decline \cite{becker2025measuring}. Indeed the "effort shifts" where GenAI often redistributes work from content creation to verification and oversight, an unrecognized form of hidden workload. So, how do we account for instances when GenAI introduce undue burden on developers well-being?

\textbf{\textit{The "Do More With Less" Trap and Invisible Oversight Labor.}} As GenAI integrates into software development pipelines, developers are required to supervise AI-generated outputs closely. This creates a new form of cognitive labor: oversight fatigue. The developer's role transitions from being an active creator to a reviewer, which may reduce autonomy and satisfaction in daily work. This supervision labor is often unmeasured and invisible, yet essential to ensure code quality and security. Current productivity metrics rarely account for the mental load and fragmented attention required to manage AI code contributions.

 The presence of GenAI can also raise an unspoken expectations and pressures on developers. If AI can generate code quickly, stakeholders may assume developers should deliver features sooner or handle more projects in parallel. This can have dire implication for the software development community and developers' well-being, especially when code propagate with errors. We are already observing this phenomenon where organizations adopting GenAI have begun escalating productivity expectations while reducing headcount and declining entry-level roles. In practice, this creates an unspoken demand for developers to \textit{"do more with less"} \cite{virtualizationreview2025technostress}. GenAI may boost productivity in one sense, but it can also impose an unsustainable fast pace of work (with questionable quality outcomes). An unsustainable work pace that demands continuous output while leaving little room for deliberation, alongside ongoing anxiety about potential failures after release, will inevitably erode developers’ well-being. Can such effects be measured? 

\textbf{\textit{Intensified Continuous Learning.}} 
Continuous learning is further intensified for software developers as GenAI rapidly reshapes workplace practices. As AI capabilities expand, developers must continuously upskill to remain effective and ensure human expertise complements AI‑driven automation in modern software development environments. In fact, AI-assisted tools such as code generation, automated testing, and intelligent debugging require developers to regularly update their technical skills and conceptual understanding \cite{amershi2019software}. Shneiderman’s human-centered AI framework \cite{shneiderman2020human} emphasizes designing AI systems that amplify human capabilities while maintaining human control, responsibility, and trustworthiness, highlighting the importance of developers’ understanding of ethical and human-centric design principles as AI becomes integral to software systems rather than an isolated component. While GenAI has the potential to lower the barriers to entry for coders, there is new demand for specialist knowledge (e.g., prompt engineering and AI code curation). This intensified continuous learning is not merely about adopting new software but also about cultivating deeper competencies that bridge traditional software expertise with emerging AI-augmented practices, enabling developers to optimize productivity while mitigating risks related to automation bias, security, and quality assurance.

In addition to individual responsibility, sustained support for continuous learning must be embedded at the organizational level to ensure effective and responsible use of AI in software development. With increasing demand on outcomes and the reorientation of what is the "team" and "a coder", developers are now required to wear many hats. In some context, developers are encouraged to avoid coding altogether, and instead see themselves as innovators/inventors \cite{stopcoding}. This changes the expectation of software developers, and demands skills that may not be typical in the regular training. Thus, organizations must conduct continuous professional training and retraining programs in enabling the development of technical skills alongside transversal competencies \cite{pelau2024can} required by GenAI . Moreover, organizations must ensure that they cultivate an appropriate culture where the use of GenAI is supported and normalized. This can be done by implementing guidelines and providing appropriate trainings \cite{hopkins2024generative} for software developers. These training programs should not only foster individual skill development but also actively promote knowledge sharing and collaboration among developers, enabling them to learn from each other and collectively build expertise in GenAI \cite{callari2025can}.

\textbf{\textit{The Social Cost of Efficiency: Erosion of Collaboration and Peer Learning.}} Although GenAI tools are often framed as personal productivity aids, their widespread and largely individual use introduces new risks to the social interaction and collaboration of software teams \cite{mohamed2025impact}. Indeed, \citet{mohamed2025impact}’s systematic literature review identifies a critical gap in research on how LLM-assisted development affects team communication and coordination, as most studies emphasize human–LLM collaboration rather than human–human collaboration with LLMs in the loop. In environments where developers increasingly turn to GenAI for assistance with debugging, documentation, or code synthesis, opportunities for organic collaboration, peer learning, and collective problem-solving may decline. Developers may rely on AI to troubleshoot rather than asking colleagues, leading to fewer moments of knowledge exchange or mentorship, reorienting the entire software development ecosystem.

This erosion of informal collaboration may be particularly detrimental for junior developers, who often learn through pair programming, asking questions, or reviewing code with more experienced peers. As AI replaces some of these interactions, there is a risk that entry-level practitioners will miss out on crucial social learning opportunities, reinforcing a sense of isolation and stalling professional growth. Moreover, as knowledge becomes increasingly fragmented—mediated by individual AI interactions rather than shared team practices—teams may struggle to maintain a common understanding of architectural decisions, design rationales, or coding conventions. In fact, most developers seek help when they lack knowledge, including knowledge to judge what is in/correct. Thus, they would be unable to evaluate what is provided by GenAI tools. With limited opportunities for peers/senior members to provide additional oversight, this reality has dire potential implication for global software quality. 

These shifts are subtle but significant. What appears to be increased efficiency at the individual level may come at the cost of weakened team cohesion and reduced collective intelligence and teams' tacit knowledge. Without intentional design and support, GenAI risks narrowing the social surface of software development, making the work faster, but lonelier.

\begin{figure*}[t]
  \centering
  \includegraphics[width=\textwidth]{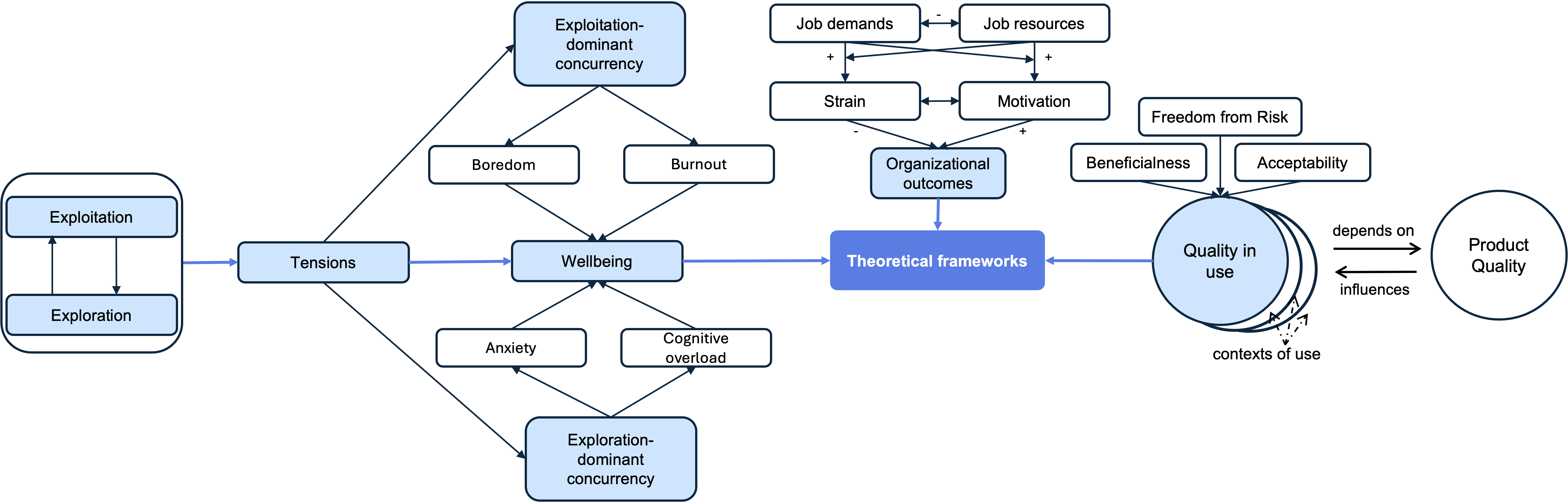}
  \caption{Proposed Theoretical Framework}
  \label{fig:wellbeing-framework}
\end{figure*}

\section{Proposed Theoretical Framework} 

In order to comprehensively examine the intended and unintended consequences of GenAI adoption on software developers’ well-being, this vision paper calls for a new, interdisciplinary approach. Such an approach should integrate three key theoretical models to deliver value, as shown in Fig \ref{fig:wellbeing-framework}: Ambidexterity theory, ISO/IEC 25019:2023 quality in use model, and Job Demand-Resource (JD-R) Model. 

Ambidexterity theory \cite{o2013organizational} explains the simultaneous existence of exploration and exploitation within organization. While Exploitation-dominant concurrency can lead to boredom and burnout, Exploration-dominant concurrency can lead to anxiety and cognitive overload. This theory is useful for explaining how the organizational requirements for concurrently exploiting and exploring GenAI tools can impact the well-being of developers. 

To evaluate user experience and the perceived effectiveness of GenAI tools, the ISO/IEC 25019:2023 provide a good basis for understanding GenAI's quality in use. According to the ISO/IEC 25019 standard quality in use encompasses the characteristics beneficialness (i.e., the overall advantage users gain from GenAI) by integrating the sub-characteritics usability, accessibility, and suitability to ensure the system supports diverse users and real work contexts. It also includes the characteristic freedom from risk, which is the extent to which Generative AI minimizes risks related to the economy, health, safety, the environment, and society, as well as the characteristic acceptability, which reflects users’ evolving experience, trust in the system, and confidence in its legal and regulatory compliance.

While both Ambidexterity theory and ISO/IEC 25019 are useful for providing insights into GenAI usage and tensions, they do not fully address the psychological and emotional impacts of GenAI integration, such as stress, workload, and engagement. To fill this gap, the JD-R model, an established framework in workplace psychology, would provide utility for exploring how job demands (e.g., increased cognitive load) and job resources (e.g., support, autonomy) influence software development practitioners’ well-being in the context of GenAI. Fig \ref{fig:wellbeing-framework} provides a strong basis for guiding our understanding of the cost of GenAI. 


Developer well-being includes stress, burnout, job satisfaction, and work–life balance, which can be evaluated using established survey scales. Cognitive load may be measured using instruments such as NASA-TLX or context-switching frequency. Other theoretical constructs such as tensions, exploitation and exploration can be measured using the established survey instruments. These operationalizations provide a concrete basis for empirical studies.

\section{
Call to Action}

We make a call to action, to bridge that gap in research focus on performance and human outcomes when examining GenAI adoption. We advocate for examining both the intended and unintended consequences of generative AI, focusing on individual experiences as well as broader social and professional implications. 

\textbf{Investigate the impact of GenAI on developers' well-being:}
Future research should explore how GenAI affects developers' mental and emotional states, including stress levels, cognitive load, job satisfaction, and work–life boundaries. This includes not only identifying harms but also recognizing where GenAI may enhance autonomy, confidence, or learning.
As GenAI tools become increasingly integrated into developers’ work, developers are required to simultaneously rely on existing tools, practices, and expertise while also exploring and learning new GenAI-based tools and workflows, \textbf{intensifying the need for continuous learning}. This dual requirement heightens the need for concurrent exploitation and exploration, as developers must maintain productivity using established approaches while adapting to rapidly evolving technologies. Such concurrency can create tensions, as competing demands on time, attention, and cognitive resources emerge, with potential implications for developer well-being. Future research should therefore examine how these exploitation–exploration tensions shape developers’ well-being outcomes, as well as the dominant mechanisms through which GenAI influences this balance. 

\textbf{Examine the social implications of GenAI adoption in the workplace:} The integration of GenAI is reshaping roles, expectations, and professional trajectories. Studies should investigate how team dynamics, mentorship structures, and career pathways are being reconfigured—especially for junior developers—within AI-augmented environments. Efforts should explore arrangements that lead to the right balance of productivity and enduring support for \textbf{collaboration and and peer learning}, ensuring good quality of GenAI use. With the human aspect of software development regularly overlooked, leading to avoidable failures, teams employing GenAI in their workflows should be explored for beneficial and risk-prone team configurations. For instance, research should explore how GenAI use affects staff-turnover and GenAI's positive and negative effects on mentoring and teams' tacit knowledge.   

\textbf{Develop strategies for the healthy and sustainable integration of GenAI:} It is essential to co-design evidence-based practices, tools, and organizational policies that prioritize developer well-being alongside performance. These strategies should account for long-term socio-technical implications, not just short-term efficiency gains, and should be rooted in diverse workplace contexts and practitioner experiences. Organizations should also adopt interventions that help manage the tension between maintaining established workflows and learning to effectively integrate emerging GenAI capabilities, not overlooking \textbf{invisible oversight labor}. Such interventions may include structured time for learning, clear guidance on when to rely on GenAI versus traditional tools, and mechanisms to mitigate cognitive overload, such as tool curation and standardized best‐practice frameworks. Supportive practices - such as peer mentoring, collaborative experimentation, and structured reflection routines - can further help developers balance exploitation and exploration without compromising well-being. 

Future research should explicitly focus on these intervention strategies, examining which are most effective in different organizational and task contexts, how they can be implemented and sustained at scale, and how they shape both performance and well-being of developers over time as GenAI technologies continue to evolve.

\balance
\bibliographystyle{ACM-Reference-Format}
\bibliography{sample-base}

\end{document}